# Realization of electron antidoping by modulating the breathing distortion in BaBiO$_3$

Hui Cao[1,2,‡], Hongli Guo[1,3,‡], Yu-Cheng Shao[4], Qixin Liu[1], Xuefei Feng[4], Qinwen Lu[1], Zhengping Wang[1], Aidi Zhao[1,5], Atsushi Fujimori[6], Yi-De Chuang[4], Hua Zhou[2*], and Xiaofang Zhai [1,5*]

[1] *Department of Chemical Physics, Department of Physics, Physics Experiment Teaching Center, and Hefei National Laboratory for Physical Sciences at the Microscale, University of Science and Technology of China, Hefei 230026, China*

[2] *X-ray Science Division, Advanced Photon Source, Argonne National Laboratory, Lemont, IL 60439, USA*

[3] *School of Physics and Technology and Key Laboratory of Artificial Micro- and Nano-Structures of Ministry of Education, Wuhan University, Wuhan 430072, China*

[4] *Advanced Light Source, Lawrence Berkeley National Laboratory, Berkeley, CA 94720 USA*

[5] *School of Physical Science and Technology, ShanghaiTech University, Shanghai 201210, China*

[6] *Department of Applied Physics, Waseda University, Okubo, Shinjuku, Tokyo 169-8555, Japan*

**Abstract**

The recent proposal of antidoping scheme breaks new ground in conceiving conversely functional materials and devices, yet the few available examples belong to the correlated electron systems. Here we demonstrate both theoretically and experimentally that the main group oxide $BaBiO_3$ is a model system for antidoping using oxygen vacancies. The first-principles calculations show that the band gap systematically increases due to the strongly enhanced Bi-O breathing distortions away from the vacancies and the annihilation of Bi $6s$/O $2p$ hybridized conduction bands near the vacancies. The spectroscopic experiments confirm the band gap increasing systematically with electron doping, with a maximal gap enhancement of ~75% when the film's stoichiometry is reduced to $BaBiO_{2.75}$. The Raman and diffraction experiments show the suppression of the overall breathing distortion. The study unambiguously demonstrates the remarkable antidoping effect in a material without strong electron correlations and underscores the importance of bond disproportionation in realizing such an effect.

**Main Text**

**Introduction**

In conventional semiconductors or insulators without strong electron correlations, the electron doping can be understood as extra electrons being injected either into the impurity donor band confined within the band gap (i.e. light doping), or into the conduction band (i.e. heavy doping) (1-3). The band gap either effectively decreases to be between the donor band and the conduction band or collapses, both leading to the conductivity enhancement. Recently, Zunger *et al*. (4) proposed the antidoping effect in a group of materials with polaron hole states or ligand hole bands, where doping electrons can cause the band gap to increase via electron filling effects. This opposite doping effect opens up new perspectives in application devices based on electron doping, such as battery cells, memory devices, electric sensors, etc. So far, the signature of antidoping effect has been observed in negative charge-transfer oxides (5-12), with strong electron correlations, such as oxygen vacancy ($O_v$) doped $SmNiO_3$ (13-16) and $NdNiO_3$ (17) and $H^+$ doped $SrCoO_{2.5}$ (18). However, the antidoping effect has not been realized in the large family of main group oxides. Unlike transition metal oxides, the main group oxides exhibit weak or negligible electron correlations since they do not have *d* electrons in the valence or conduction band. Thus it is important to investigate the possibility of antidoping effect in main group oxides and further unravel alternative mechanisms that are beyond the electronic effects.

The main group oxide $BaBiO_3$ is a negative charge-transfer insulator (19-21) and thus a model system for investigating the antidoping effect (4). Bi in $BaBiO_3$ has a formal 4+

valence ($Bi^{4+}$) with a $5d^{10}6s^1$ electronic configuration, and this material has drawn long-lasting interests in the hole doping sides due to the superconductivity (22-26). On the contrary, the electron doping side has rarely been studied (27). The stoichiometric $BaBiO_3$ is an insulator with a direct band gap near 2 eV (22,28-30) and the insulating nature has been attributed to the strong hybridization between Bi and O orbitals and their charge and bond disproportionation (19,21,23,31,32). Consequently, the bismuth 6s and oxygen 2p orbital both have significant contributions to the conduction band nearest to the Fermi level (20,21), which is different from pure ligand hole states discussed in ref. (4). The option of doping oxygen vacancies ($O_v$s) to realize electron doping has the advantage in clearly revealing the evolution of band structure without complications from the dopant bands. In addition, $O_v$s can be driven by electric field which promises facile applications (16).

Here we perform both theoretical calculations and comprehensive experiments to investigate the electron antidoping effect in $BaBiO_{3-\delta}$ ($0 \leq \delta \leq 0.25$) thin films. It is theoretically found that the insulating gap increases with increasing $O_v$ concentration due to the strongly enhanced Bi-O breathing distortions away from the $O_v$s and the annihilation of Bi 6s/O 2p hybridized conduction bands near the $O_v$s. $BaBiO_3$ and $BaBiO_{3-\delta}$ thin films were experimentally fabricated with Bi valence decreasing from 4+ to about 3.5+. The X-ray absorption spectroscopy (XAS) and X-ray emission spectroscopy (XES) experiments demonstrate that the band gap systematically increases with increasing $O_v$ concentrations. Furthermore, the Raman spectroscopy and superstructure-half-order X-ray diffraction (XRD) measurements reveal the lattice breathing distortion being partially reduced, which agree to the theory. Therefore, our study demonstrates a unique mechanism for realizing electron antidoping in BaBiO$_3$, which might inspire new prospects in controlling the electronic structures of main group oxides and related materials.

**Results**

The density functional theory (DFT) calculations using hybrid xc-functional (HSE06) were performed to elucidate the electronic structure change from the stoichiometric BaBiO$_3$ to O$_v$ doped BaBiO$_{3-\delta}$ ($\delta$ = 0.25, 0.125). The calculated monoclinic BaBiO$_3$ unit cell contains two non-equivalent Bi atoms labeled as Bi1 and Bi2 as shown in the left panel of Fig. 1*A*. The optimized Bi1-O and Bi2-O bond lengths are 2.28 Å and 2.16 Å respectively, and the tilting instability is observed with a tilting angle $t$ = 9.49°. The difference in the bond length is $b$ = 0.12 Å, which represents the breathing distortion and agrees well to previous experiments (33,34) and calculations (35,36). When two O$_v$s are introduced into to the 2x2x1 supercell at various locations (Supplementary Fig. 1) to keep the stoichiometry at BaBiO$_{2.75}$, the most stable configuration is shown in the right panel of Fig. 1a. The O$_v$ resides in between Bi3 and Bi4, while Bi1 and Bi2 are away from O$_v$s. The BiO$_5$ tetrahedrons around Bi3 and Bi4 have similar bond lengths of ~ 2.26 Å. However, an enhanced breathing distortion on the Bi1 and Bi2 sites is observed and the long and short bond lengths are 2.34 Å and 2.18 Å respectively, which leads to an increased bond length difference $b$ increased to 0.16 Å.

The calculated band structures of BaBiO$_3$ and BaBiO$_{2.75}$ are shown in Fig. 1b and Fig. 1c,

respectively, which demonstrates that the band gap increases from 0.44 eV in BaBiO$_3$ to 0.76 eV in BaBiO$_{2.75}$. In Figs. 1d and 1e the calculated DOS of different orbitals of BaBiO$_3$ and BaBiO$_{2.75}$ are shown respectively. The difference charge density map of BaBiO$_{2.75}$ is also calculated to demonstrate the electron transfer between bismuth and oxygen (Supplementary Fig. 3). For BaBiO$_3$, the breathing distortion $b$ and the tilting angel $t$ are well correlated to the charge disproportion between two Bi atoms, which can be seen in the 6$s$ projected DOS (pDOS) in the top panel of Fig. 1d. The Bi1 6$s$ orbital has more spectral weights below the Fermi level whereas Bi2 6$s$ orbital has more weights above it. There exists some back transfer of 6$p$ charge to compensate the 6s charge disproportionation, thus the 6$p$ orbitals exhibit an opposite trend of disproportionation compared to the 6s orbitals (middle panel of Fig. 1d). It results in a charge order of Bi$^{(3+x)+}$ and Bi$^{(5-x)+}$ (0 < $x$ < 1) in the stoichiometric BaBiO$_3$ (36). In addition, we see the nearly identical DOS profile for Bi2 6$s$ and O 2$p$ orbitals in the bottom panel of Fig. 1d, suggesting the strong hybridization between them. For BaBiO$_{2.75}$, the Bi1 and Bi2 ions have the stronger inequivalent projected 6$s$ DOS and larger band gap than those in the stoichiometric BaBiO$_3$ (top panel in Fig. 1d). It indicates that Bi1 and Bi2 are transitioned into Bi$^{5-x'}$ and Bi$^{3+x'}$ with a much stronger charge disproportionation. As a result, the enhanced breathing distortion on the Bi1 and Bi2 sites away from O$_v$s is key to the enhancement of band gap. Our calculations also show that Bi3 and Bi4 ions exhibit strongly suppressed charge disproportionation (middle panel of Fig. 1e). Compared to the undoped counterpart, the DOS in the 6$s$ conduction band of Bi4 vanishes, along with the strongly suppressed DOS in conduction bands of the surrounding O $p$ orbitals. Our finding is different from the previous study (4) since in the latter case, the band gap enhancement is mainly through the electron filling effect without the structural response. While in the current case, it is the response of electronic structure to the structural instability under $O_v$ doping. Since the band gap measurements are usually performed at elevated temperatures and the DFT calculations describe the ground state at 0 K, we perform *ab initio* Molecular Dynamics at 300 K in order to take the electron-phonon coupling into account. The obtained energy evolutions of electronic states in Supplementary Fig. 4, which confirms the trend of band gap enhancement in $BaBiO_3$, $BaBiO_{2.875,}$ and $BaBiO_{2.75}$ at elevated temperatures. We also calculated the phonons dispersion spectra and found that the structures remain stable when $O_v$s are introduced (Supplementary Fig. 5).

To experimentally test the theoretical calculation about this novel antidoping scheme, we fabricated $BaBiO_3$ and $BaBiO_{3-\delta}$ thin films using the high-pressure reflective high energy electron diffraction (RHEED) assisted pulsed laser deposition (PLD). The samples were grown and *in-situ* annealed in the PLD chamber to obtain different levels of $O_v$ concentration. More details about the sample fabrication can be found in Methods section. Fig. 2a shows the first derivative (d$I$/d$E$, $I$ is the intensity and $E$ is the energy) of the X-ray absorption near-edge spectra (XANES) reported in Supplementary Fig. 6a. The observed peak A, B, and C represent transitions from Bi $2p_{3/2}$ to $6s$, $6d$ $t_{2g}$ and $6d$ $e_g$ orbitals, respectively (37, 38). It can be seen that as Bi valence changes from 3+ ($6s^2$) to 5+ ($6s^0$), peak A moves to the higher energy while peak C moves to the lower energy. The opposite trends in the energy shifts of peak A and C are due to the fact that the $6s$ state becomes partially empty and $6d$ state remains empty. In comparison, the shift of peak B is negligible. Due to the resolution limit, we compared the spectral lineshape of d$I$/d$E$ and the Bi valence was determined from the integrated area of the three peaks shown in Fig. 2a after proper background subtraction (Supplementary Fig. 6b). By assuming a linear relation between the oxygen stoichiometry and the integrated area (via Bi valence quantification), the oxygen stoichiometry of different samples can be derived and is shown to the right axis of Fig. 2b, where the highest $O_v$ concentration corresponds to the BaBiO$_{2.75}$ phase. This method is justified since these three peaks tend to collapse together when Bi valence changes toward 3+ as a result of the filling of 6$s$ state and the reduced crystal-field splitting between 6$d$ $t_{2g}$ and $e_g$ states. On the other hand, when Bi valence changes toward 5+, these peaks become more separated. Furthermore, we measured the in-plane and out-of-plane lattice constants of the BaBiO$_{3-\delta}$ films (Supplementary Fig. 8b) and found that the obtained lattice constants agree well with bulk BaBiO$_{3-\delta}$ samples with similar $\delta$ (39).

We measured the optical conductivity $\sigma(\omega)$ of the BaBiO$_3$ and BaBiO$_{3-\delta}$ films using spectroscopic ellipsometry. In Fig. 2c, we observed a strong absorption peak near 2.1 eV in BaBiO$_{3.05}$ and BaBiO$_{2.95}$ that have high oxygen concentrations. As oxygen atoms are stripped out of the film, the absorption peak shifts to higher energies and is about 2.2 eV in BaBiO$_{2.75}$. The optical conductivity over a large spectral range is shown in Supplementary Fig. 9b which indicates the change of the background signal underneath the ~2 eV peak is not responsible for the peak shifting. Another striking observation is that the peak height drops quickly as the amount of $O_v$s increases. The absorption peak near 2 eV has been attributed to the gap opening due to the bond disproportionation

(19-20, 28-30). Because the ellipsometry measurements were limited on the low energy side (~ 0.7 eV), the onsets of the absorption are not observed. We compare the photon energy $E(\sigma_0)$ at which the optical conductivities all have a very small value $\sigma_0 = 23$ $\Omega^{-1}cm^{-1}$ (Supplementary Fig. 9a). As shown in Fig. 2d, both the absorption peak position and the $E(\sigma_0)$ increase with the amount of $O_V$s. Additionally, we measured the ellipsometry of another set of duplicated samples which gave qualitatively similar results (Supplementary Fig. 10).

In order to directly quantify the band gaps of the $BaBiO_3$ and $BaBiO_{3-\delta}$ thin films, we employed the X-ray absorption (XAS) and emission spectroscopy (XES) (40) to probe both the unoccupied and occupied DOS (the sketch of XAS and XES is shown in Fig. 3a). Both measurements were done using the fluorescence yield method. The first derivatives of XAS and XES spectra are shown in Fig. 3b. The first derivative enhances the shallow peak (from the film) that would otherwise be masked by much stronger features (from the substrate) further away from the Fermi level. We obtain the band gap through fitting the first-derivative of XES and XAS spectra with Gaussian functions (41-43) and the fitting errors in Fig. 3b are marked as band gap errors. The band gaps measured by XAS and XES are summarized in Fig. 3a. From the fitting results, it can be seen that the band gap systematically increases from ~ 0.59 eV to ~1.04 eV when the $O_v$ concentration is increased. The maximal determined enhancement is about 75%, well matched by the DFT prediction. Thus, the systematic band gap enhancement trend with increased $O_v$ concentration is identified, which experimentally confirms the antidoping effect.

Furthermore, we performed Raman spectroscopy and synchrotron half-order XRD measurements to investigate the breathing distortions in BaBiO$_{3-\delta}$ ($0 \leq \delta \leq 0.25$) films. Experimental details can be found in Supplementary section III. The sketch of Bi-O breathing distortion is shown in Fig. 4a. First, the Raman experiments qualitatively show the change of the breathing distortion. Previous studies have confirmed that the Raman mode at $\approx 570$ cm$^{-1}$ corresponds to the breathing distortion of stoichiometric BaBiO$_3$ (30,44). The current results are shown in Fig. 4b, which demonstrate that the amplitude of the Raman mode at 570 cm$^{-1}$ is weakened as $\delta$ increases. Thus the overall breathing distortion is expected to be weakened in films with large $\delta$. Secondly, the synchrotron half-order XRD intensity profile is analyzed to quantitatively determine the oxygen-octahedral-rotation (OOR) (45-47) and breathing distortion (Supplementary section III). In Fig. 4c, the rotations around the pseudocubic $a$, $b$, $c$ axes with OOR angles of $\alpha$, $\beta$, $\gamma$ are schematically shown. The three rotation angles and the breathing distortion $\Delta$ (the percentage variation of the Bi-O bond length) are shown in Fig. 4d. It is observed that in lightly oxygen-deficient samples (small $\delta$), the $\alpha/\beta$ is much larger than $\gamma$, which is consistent with the orthorhombic symmetry. While in highly deficient samples (large $\delta$), the $\alpha/\beta$ is similar to $\gamma$ which is in line with a more symmetric structure revealed by the lattice constant analysis (Supplementary section III, 39). The experimentally derived breathing distortion $\Delta_E$ is compared to the theoretical $\Delta_T$. It is obvious that both $\Delta_T$ and $\Delta_E$ are much larger in the highly oxidized samples than the poorly oxidized samples. Therefore the above experimental findings support the modulated lattice breathing distortion as the underlying reason of the antidoping effect.

**Discussion**

The antidoping effect in BaBiO$_3$ found in the current study is driven by the response of the structural instability (breathing distortion) to the O$_v$ doping, which is indeed different from other antidoping cases that involve transition metal elements. In those cases, the electron doping is related to large Coulomb interaction and the strong electron correlation has been used to explain the observed band gap enhancement in electron doped nickelate oxides (13-15, 17). Theoretically the doping level needs to reach 1 $e$ per transition metal ion in order to activate the electron correlation and enlarge the band gap (4, 16). While in the current study, the partially enhanced breathing distortion on the O$_v$-free Bi1-O and Bi2-O sites (see Fig. 1) increases the gap at doping levels much less than 1 $e$/Bi. Moreover, in previous theoretical proposals of antidoping binary oxides (MgO, ZnO, etc.), defects such as metal vacancies are prerequisites to create polaronic hole states for the electron antidoping to take place (4). While in antidoping BaBiO$_3$, the antidoping is initiated in the undoped phase. Additionally, the BaBiO$_3$ has very strong covalent bonds between Bi and O atoms. As a result, the oxygen hole states are extended Bloch bands which are different from those in doped binary oxides. Therefore the study of BaBiO$_3$ opens up a new route to antidoping main group materials with the underlying mechanism different from previous ones.

**Conclusion**

In summary, we demonstrate the antidoping effect induced in a main group oxide via the strong lattice response under the O$_v$ doping using both DFT theory calculations and comprehensive experiments. The theory calculations show that the band gap systematically increases with $O_v$ doping due to the enhanced bond disproportionation in Bi-O bonds away from $O_v$s and the suppressed bond disproportionation in Bi-O bonds next to $O_v$s. Experimentally, the XAS and XES measurements confirm that the band gap systematically increases with $O_v$ doping, with a maximal gap enhancement of ~75% in $BaBiO_{2.75}$. The Raman and half order XRD experiments demonstrate the weakening of the overall breathing distortion in agreement with theory. Thus the successful antidoping of $BaBiO_3$ has been demonstrated with a root mechanism associated with the strong electron-lattice interaction. The current study manifests the extraordinary electron antidoping phenomenon in a main group oxide and opens up a new avenue for tailoring their electronic properties.

## Methods

**First-Principles DFT Calculation.**

The electronic properties of bulk BaBiO$_3$ and those with oxygen vacancies (O$_v$s) were elucidated using first-principles calculations, as implemented in VASP. The electron-ion interactions are described by the projector-augmented wave method. For structure optimization, we used the generalized gradient approximation (GGA) with PBE functional, for static electronic structure calculations, the HSE06 functional was used (48). In our calculation, all of the atomic positions are relaxed until the force on each atom is less than 0.01 eV/Å, and all of the self-consistent electronic calculations are converged to 10$^{-5}$ eV per cell. The primitive monoclinic unit cell with $a = b = 6.17$ Å and $c = 6.23$ Å, $\alpha = 60°$, $\beta = \gamma = 120°$, C2/m (12) space group, containing 10 atoms (i.e. two BaBiO$_3$) was used for bulk calculation (as shown in the left side of Fig. 1a). The fractional coordinates of the atoms are listed in Supplementary Table 1. A 9x9x9 k-points mesh grid and an energy cutoff of 400 eV were used. In order to simulate BaBiO$_{2.75}$, a 1x2x1 supercell with one O$_v$ was built (shown in the right side of Fig. 1a). For the supercell calculations, 9x5x9 k-points grid was sampled. Since it is well known that the PBE functional underestimates the band gap due to the self-interaction error (49), our PBE result for the pristine bulk showing the metallic property is consistent with previous studies (36). Thus the HSE06 functional is taken into account in order to produce the reasonable electronic structure for BaBiO$_3$.

**BaBiO$_{3-\delta}$ ($0 \leq \delta \leq 0.25$) thin films growth**

All thin films were grown using High-Pressure Reflective High-Energy Electron

Diffraction Assisted Pulsed-Laser Deposition system. All BaBiO$_{3-\delta}$ ($0 \leq \delta \leq 0.25$) were grown on SrTiO$_3$ (001) single crystalline substrates from a stoichiometric BaBiO$_3$ target at the substrate temperature of 700 °C with an oxygen pressure of 20 Pa. The laser energy density was 1.6 J/cm$^2$ and the pulse frequency was 1 Hz. After growth, the partial pressure of oxygen and the time for post annealing were controlled to obtain films with different oxygen stoichiometry. The stoichiometric BaBiO$_3$ films were obtained by annealing the as-grown films with 400 kPa oxygen pressure for 30 minutes. The O$_v$ doped films were obtained by annealing the as-grown films for 5, 10 and 15 minutes with a low oxygen pressure of 0.5 Pa, respectively. The thickness of films was precisely controlled by RHEED oscillations and then confirmed by X-ray reflectivity measurements. We fabricated two other groups of samples and repeated the structural and spectroscopic measurements which confirmed the reproducibility.

**X-ray absorption and emission spectroscopy and half-order X-ray diffraction**

Bi $L_3$-edge XANES measurements were performed at 12-BM-B at Advanced Photon Source, using a Si (311) monochromator with an energy range of 10-40 keV and the energy resolution $\Delta E/E$ is $2 \times 10^{-5}$. The soft X-ray absorption spectroscopy (XAS) and soft X-ray emission spectroscopy (XES) measurements were conducted in the qRIXS endstation at beamline 8.0.1 at the Advanced Light Source (ALS), Lawrence Berkeley National Laboratory. All XAS spectra in the total fluorescence yield (TFY) mode were carried out at room temperature. The half order X-ray diffractions were carried out on 1W1A (10 keV) at Beijing Synchrotron Radiation Facility and 33-BM-C (16 keV) at Advanced Photon Source in Argonne National Laboratory. More details of the XAS, XES experiments are described in the Supplementary section II and half-order XRD and Raman experiments in Supplementary section II and III.

**Acknowledgments**

The authors acknowledge Profs. Yi Luo, Bing Wang, Xuefeng Cui and Rui Peng for helpful discussions and experiments. The work was supported by the National Key Research and Development Program of China (Grants No. 2016YFA0401004), National Science Foundation of China (52072244), China Postdoctoral Science Foundation (Grant No. 2017M622498 and 2018T110789), and KAKENHI from JSPS (19K03741). Calculations are performed at the Supercomputer Center of Wuhan University and Environmental Molecular Sciences Laboratory at the PNNL. This research used resources of the Advanced Photon Source (APS), a U.S. Department of Energy (DOE) Office of Science User Facility operated for the DOE Office of Science by Argonne National Laboratory under Contract No. DE-AC02-06CH11357. The synchrotron XRD and XANES measurements were carried out at Sector 33BM, 12ID-D, 12BM-B of the Advanced Photon Source and 1W1B of the Beijing Synchrotron Radiation Facility. This research used resources of the Advanced Light Source, a DOE Office of Science User Facility under contract No. DE-AC02-05CH11231. Some XAS experiments were done at BL12B-a at the National Synchrotron Radiation Laboratory.

[‡] H. C. and H.G. contributed equally to this work.

[*] To whom correspondence may be addressed. Email: zhaixf@shanghaitech.edu.cn or hzhou@anl.gov

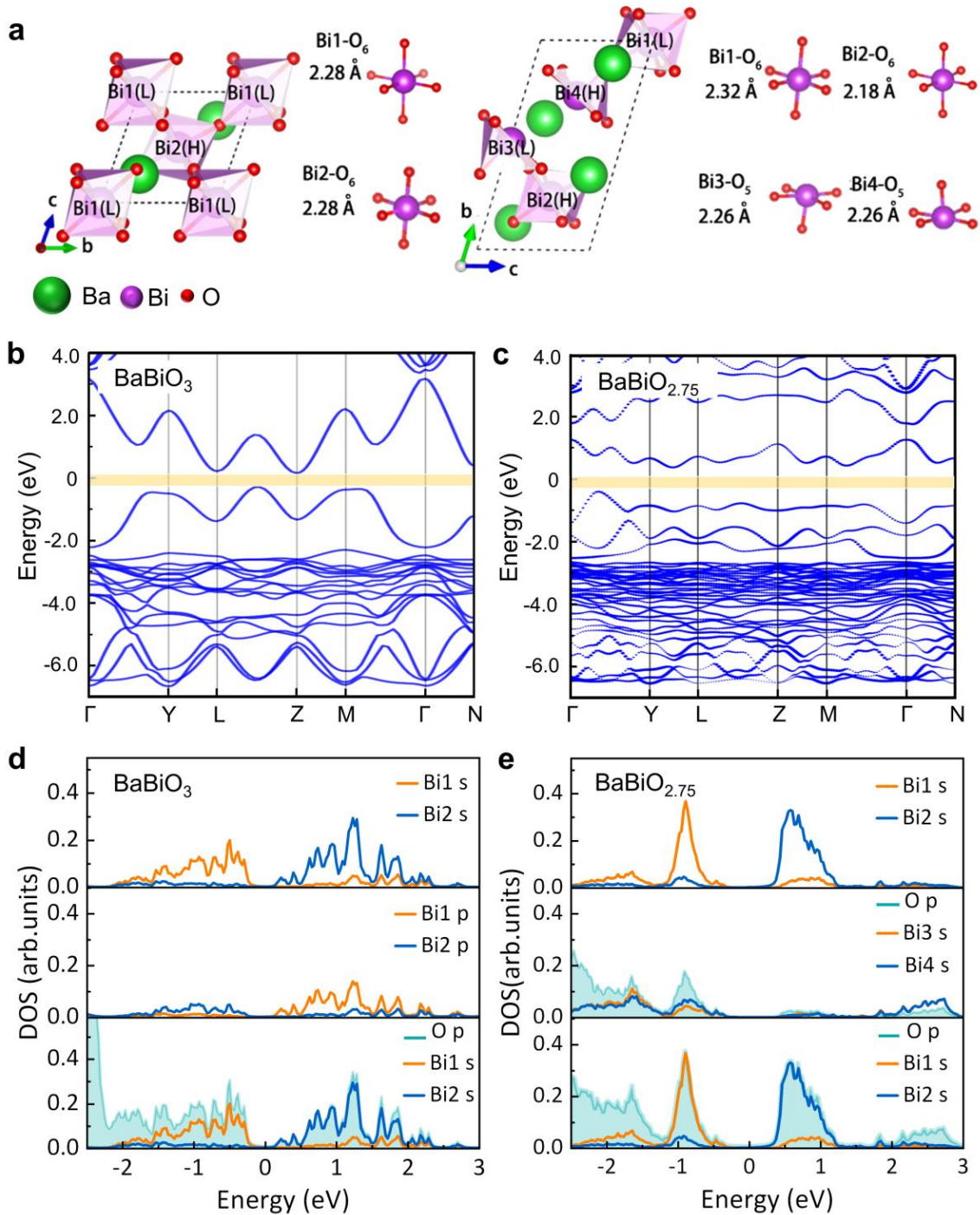

**Fig. 1.** (a) Illustrations of atomic structures before (left) and after (right) $O_v$ being introduced. Bi3 and Bi4 are identical to Bi1 and Bi2 before the $O_v$ introduction. (b) The band structures of $BaBiO_3$ and (c) $BaBiO_{2.75}$; the yellow shade indicates the band gap of $BaBiO_3$. (d) (top panel) The *s* projected DOS of Bi1 and Bi2 ions in $BaBiO_3$, (middle

panel) the *p* projected DOS of Bi1 and Bi2 ions, and (bottom panel) the overlay of Bi1/Bi2 *s* and O *p* orbitals. (e) (top panel) The *s* projected DOS of Bi1 and Bi2 in BaBiO$_{2.75}$, (middle panel) the *s* projected DOS of Bi3 and Bi4 and surrounded O *p* orbitals, and (bottom panel) the overlay of Bi1/Bi2 *s* and surrounded O *p* orbitals.

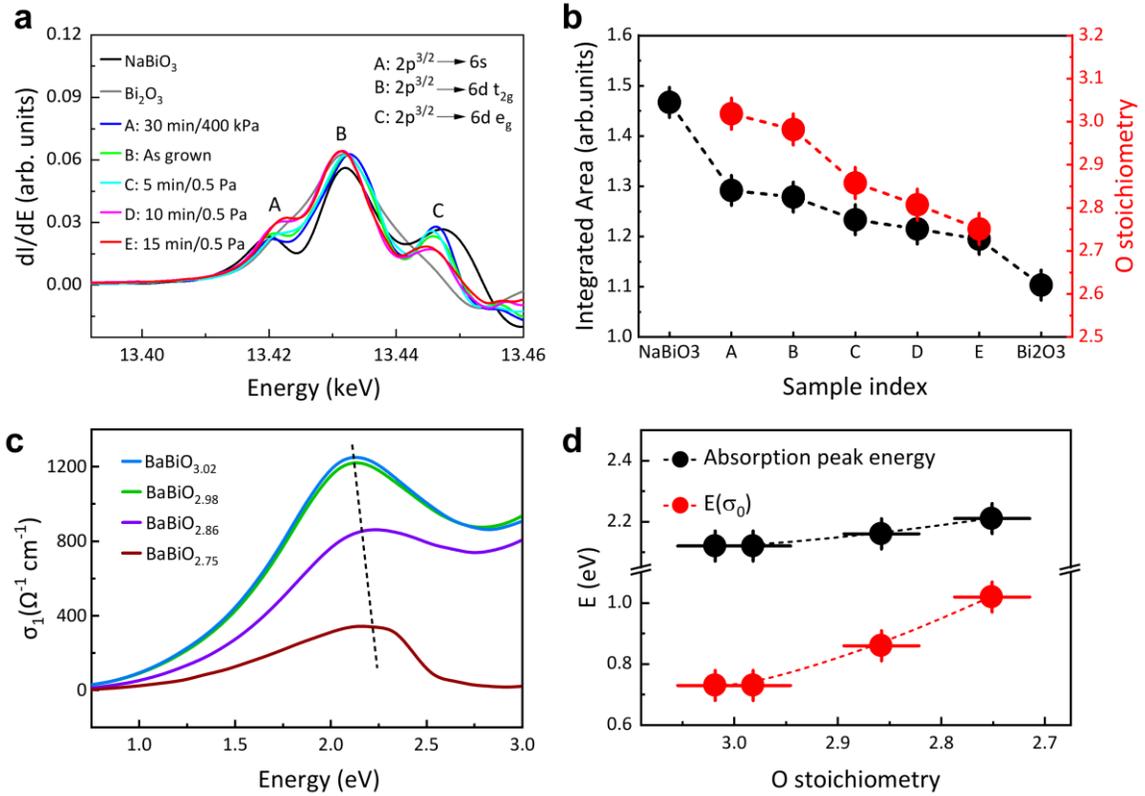

**Fig. 2.** (a) The first derivative of Bi $L_3$-edge XANES spectra of the BaBiO$_{3-\delta}$ ($0 \leq \delta \leq 0.25$) films and two reference samples. (b) The correlation between the integrated area of derivative XANES spectral and the oxygen stoichiometry of BaBiO$_{3-\delta}$ ($0 \leq \delta \leq 0.25$) films and two reference samples. (c) The real part of optical conductivity $\sigma_1$ measured via spectroscopic ellipsometry. The dashed line indicates the absorption maximums. (d) The absorption peak energy and the photon energy corresponding to a small optical conductivity $\sigma_0 = 23$ $\Omega^{-1}$cm$^{-1}$ of four films. All dashed lines are guides for eye.

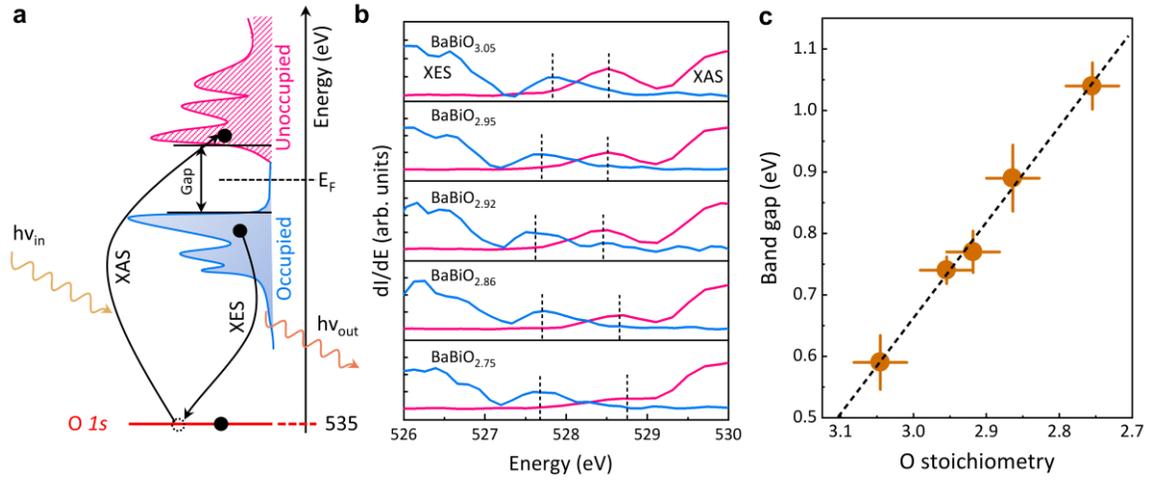

**Fig. 3.** (a) Sketch of XAS and XES experiments; (b) The first derivative of XAS (pink) and XES (blue) spectra with determined band gap positions. (c) Band gaps of $BaBiO_{3-\delta}$ ($0 \leq \delta \leq 0.25$) thin films; the short-dashed line is a guide for eye.

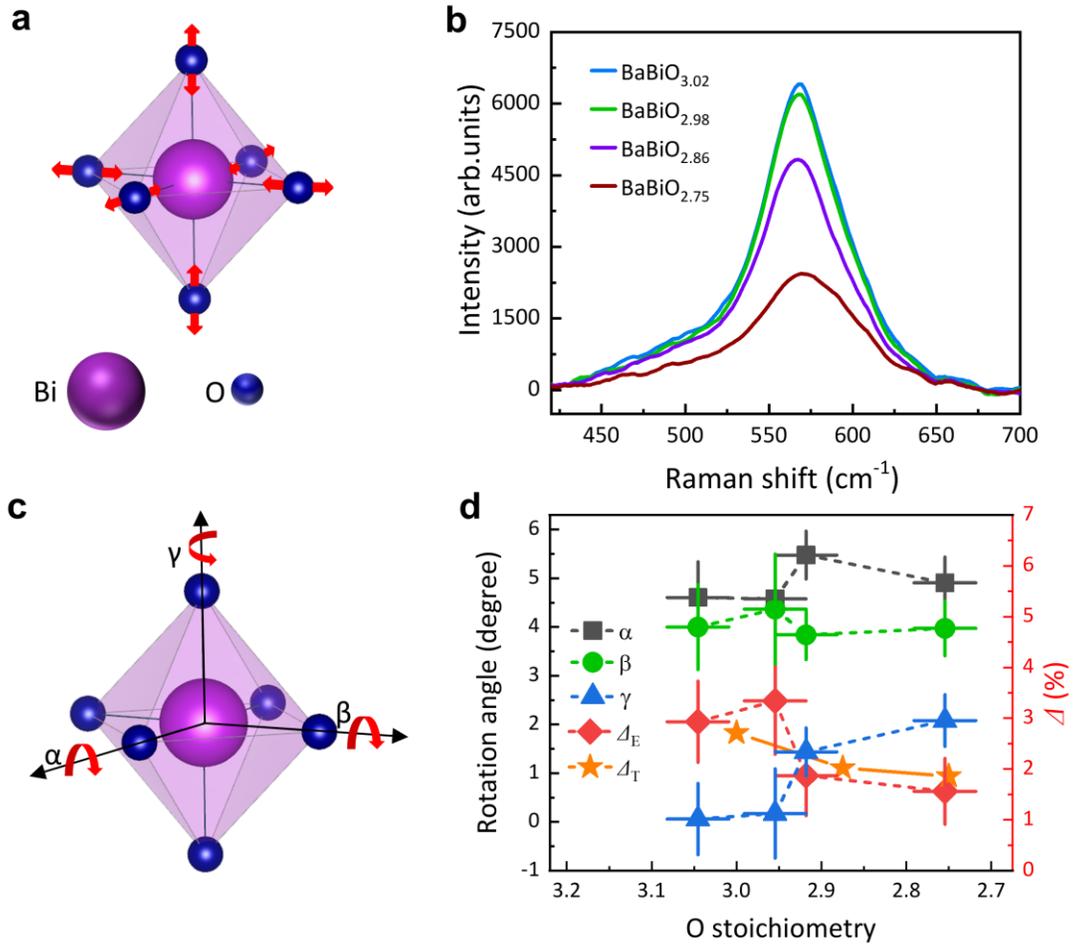

**Fig. 4.** (a) Sketch of the Bi-O bond breathing distortion; (b) Raman spectra for BaBiO$_{3-\delta}$ films measured at 300 K under a 514.5 nm-Ar-laser excitation. The spectral features show a mode at ∼570 cm$^{-1}$, which results from the oxygen breathing phonon (see Fig. 4(a)). (c) Sketch of oxygen octahedral rotation (OOR) geometry; (d) The breathing distortion amplitude measured by experiments ($\Delta_E$) and calculated by theories ($\Delta_T$) and OOR angles ($\alpha$, $\beta$, $\gamma$) of the BaBiO$_{3-\delta}$ films.

# Supplementary Information: Realization of electron antidoping by modulating the breathing distortion in BaBiO$_3$

Hui Cao[1,2,‡], Hongli Guo[1,3,‡], Yu-Cheng Shao[4], Qixin Liu[1], Xuefei Feng[4], Qinwen Lu[1], Zhengping Wang[1], Aidi Zhao[1,5], Atsushi Fujimori[6], Yi-De Chuang[4], Hua Zhou[2*], and Xiaofang Zhai [1,5*]

[1] *Department of Chemical Physics, Department of Physics, Physics Experiment Teaching Center, and Hefei National Laboratory for Physical Sciences at the Microscale, University of Science and Technology of China, Hefei 230026, China*

[2] *X-ray Science Division, Advanced Photon Source, Argonne National Laboratory, Lemont, IL 60439, USA*

[3] *School of Physics and Technology and Key Laboratory of Artificial Micro- and Nano-Structures of Ministry of Education, Wuhan University, Wuhan 430072, China*

[4] *Advanced Light Source, Lawrence Berkeley National Laboratory, Berkeley, CA 94720 USA*

[5] *School of Physical Science and Technology, ShanghaiTech University, Shanghai 201210, China*

[6] *Department of Applied Physics, Waseda University, Okubo, Shinjuku, Tokyo 169-8555, Japan*

## Section I. XRD characterization of lattice constants

The reciprocal space map (RSM) measurements were done to detect the in-plane lattice constants of the $BaBiO_3$ and $BaBiO_{3-\delta}$ films, with the results shown in Supplementary Fig. 7. We measured the lattice structure of the $BaBiO_3$ and $BaBiO_{3-\delta}$ films using high-resolution x-ray diffraction (XRD). The monochromatic X-rays were provided by Cu $K_{\alpha 1}$ ($\lambda$ = 0.154056 nm and $\Delta\lambda/\lambda$ ~23 ppm), the beam divergence was 12 arc sec. An additional monochromator with a two-bounce Ge (220) channel-cut analyzer was placed in front of a proportional counter in the diffracted beam path with the same beam divergence. For polar angle ($\theta$) and tilt angle ($\varphi$), the angular precision and reproducibility of the diffractometer are 0.0001° and 0.0003°, respectively. The (0 0 $L$) specular reflection of the $BaBiO_3$ and $BaBiO_{3-\delta}$ films around $L$=2 exhibits clear thickness oscillation fringes (see Supplementary Fig. 8a), demonstrating the high quality of these thin film samples. Both in-plane and out-of-plane lattice constants of these films are plotted in Supplementary Fig. 8b. As the amount of $O_v$s increases, the out-of-plane lattice constant $c$ increases from 4.34(0) Å to 4.35(8) Å while the in-plane lattice constant $a$ decreases from ~4.38 Å to 4.35(3) Å. Overall, the lattice transforms from a slightly tetragonal ($a > c$) structure to a pseudocubic ($a \approx c$) structure. Such a structural transition and the almost conserved unit cell volume agree with the bulk study of $BaBiO_{3-\delta}$ in the range of $0 \leq \delta \leq 0.25$ (1).

## Section II. Details of X-ray absorption and emission spectroscopy measurements

For the O K-edge XAS and XES measurements, all thin-film samples were attached to the sample holders with carbon tapes and transferred from airside into the high vacuum experimental chamber (pressure better than $8\times10^{-9}$ torr) through a load-lock system. All measurements were carried out at room temperature with the beamline energy resolution set to ~ 0.4 eV at O K-edge and π-polarization in the horizontal scattering plane. The sample surface was at 30º grazing incidence angle and a GaAsP photodiode located at 150º emission angle relative to the incident x-ray beam was used to record the XAS spectra. This photodiode, with an angular acceptance of 30 mrad (both vertical and horizontal), has a frontal Al window to block out the ambient visible light and photoelectrons. For XES, the excitation photon energy was set to 570 eV, far above the O absorption edge. A modular X-ray spectrometer at 140º emission angle was used to record the XES spectra (2). The combined energy resolution (~ 0.4 eV, beamline plus spectrometer) and the calibration for XES spectra (converting the detector pixel number to emission energy) were determined using the elastic line from a carbon tape next to the samples. This method ensures the proper energy alignment between XAS and XES to be better than 0.05 eV.

**Section III. Details of Raman spectroscopy, XRD half-order Bragg peak measurement and quantification of oxygen octahedral rotations**

Raman spectra were measured using a laser confocal spectrometer with an excitation wavelength of 514 nm and a magnification level of 100. The spectra were obtained with a laser spot size of about 1 μm, an illumination power less than 20 mW, and an integration time of 10 sec. The spectral resolution for the incoming 514 nm light is 0.65 cm$^{-1}$.

It is known that the oxygen-octahedral-rotation (OOR) induces a unique XRD intensity profile of half-order Bragg peaks, which can be utilized to determine the rotation angles

(3-5). For detailed XRD half-order-peak investigations, the XRD measurements used hard X-Rays source of sector 33BM of the Advanced photon Source at room temperature. To rule out the fluorescence signal of the substrate, the high energy (E = 16 keV) of X-Ray has been chosen. The geometric corrections and background subtractions were done for all data. Then we analyzed the data using the previous fitting procedure (3-6). In $BaBiO_3$, in addition to the OOR, the breathing distortion also doubles the unit cells along each of the pseudocubic axis, resulting in a supercell with 24 oxygen atoms which is similar to the OOR effect. In Supplementary Fig. 11, the $BiO_6$ octahedra are shown to collapse and expand alternately along with the three pseudocubic crystallographic directions. Here the positions of the oxygen atoms are first individually calculated from the rotation matrix as in previous literature (3-5). Then the oxygen position is further modulated by the bond length correction due to the breathing distortion. The rotation angles $\alpha$, $\beta$ and $\gamma$ are along the pseudocubic $a$, $b$ and $c$ axis, respectively. The breathing distortion $\Delta$ defines the bond variation, with the collapsed bond and elongated bond length to be $(1-\Delta)L$ and $(1+\Delta)L$ respectively. $L$ is the average bond length of Bi-O but is not appearing in the half order peak profile calculation (3-5). For example, in Fig. S11, the Bi-O bond lengths in the #1, #4, #6 and #7 octahedra are $(1+\Delta)L$, and the Bi-O bond lengths in the #2, #3, #5 and #8 octahedra are $(1-\Delta)L$. By considering both the rotations and the bond length variations, the positions of 24 oxygen atoms are all defined by the four independent variables of $\alpha$, $\beta$, $\gamma$ and $\Delta$. The typical fitting results of $BaBiO_{2.92}$ are shown in Supplementary Fig. 12.

**Supplementary Table 1:** The positions of atoms in primitive cell of monoclinic BaBiO$_3$ unit cell.

| Atom label | Position (x, y, z) |
|---|---|
| Ba1 | (0.249629, 0.750371, 0.741814) |
| Ba2 | (0.750371, 0.249629, 0.258186) |
| Bi1 | (0.000000, 0.000000, 0.000000) |
| Bi2 | (0.500000, 0.500000, 0.500000) |
| O1 | (0.702278, 0.785292, 0.699562) |
| O2 | (0.785292, 0.702278, 0.300438) |
| O3 | (0.297722, 0.214708, 0.300438) |
| O4 | (0.214708, 0.297722, 0.699562) |
| O5 | (0.258392, 0.741608, 0.183981) |
| O6 | (0.741608, 0.258392, 0.816019) |

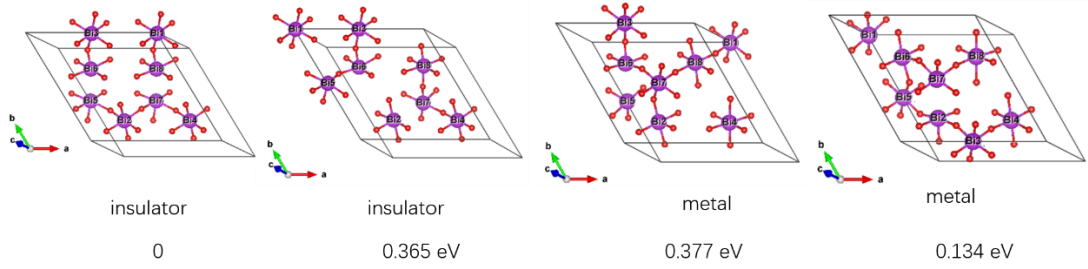

**Supplementary Figure 1** Four different configurations with different $O_v$ locations in $BaBiO_{2.75}$; the most stable one is identical to that in the main text.

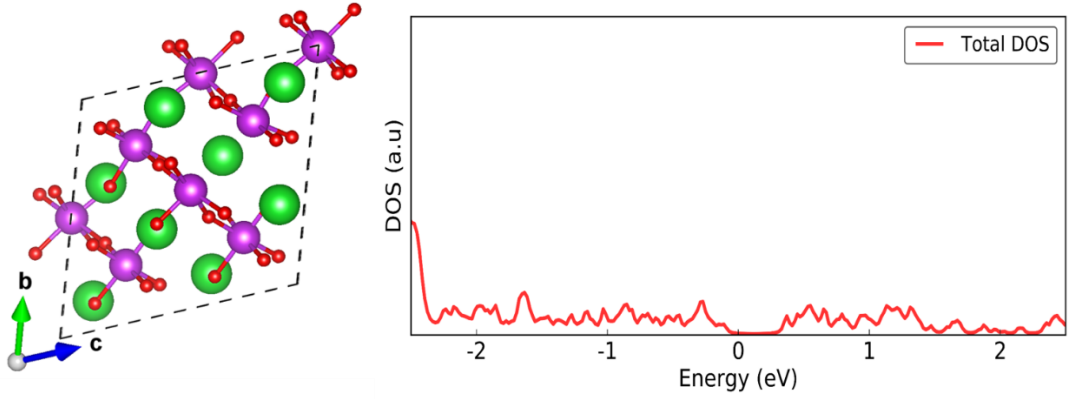

**Supplementary Figure 2** Left: configuration of a 2x2x2 supercell with 2 $O_v$s (BaBiO$_{2.875}$); Right: the DOS of BaBiO$_{2.875}$, still showing an insulating gap.

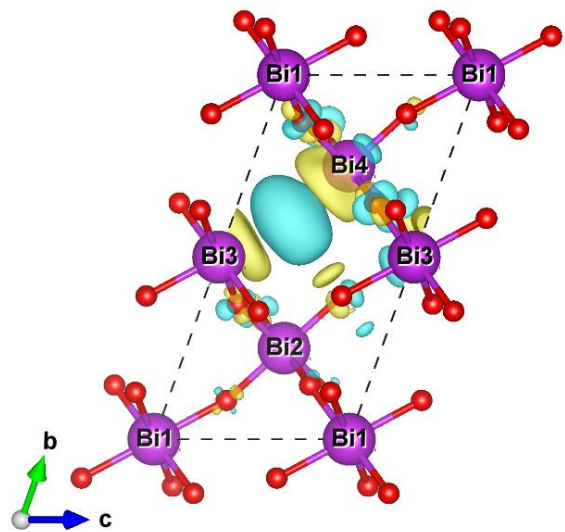

**Supplementary Figure 3** The difference charge density map of BaBiO$_{2.75}$, the yellow iso-surface represents the charge accumulation, and blue iso-surface represents the charge depletion. From the difference charge density, we can see that the electrons are transferred from the oxygen vacancy site to both Bi3 and Bi4 atoms; however, Bi4 atom gets more electrons than Bi3. In addition, charge transfer also happens at the oxygen sites.

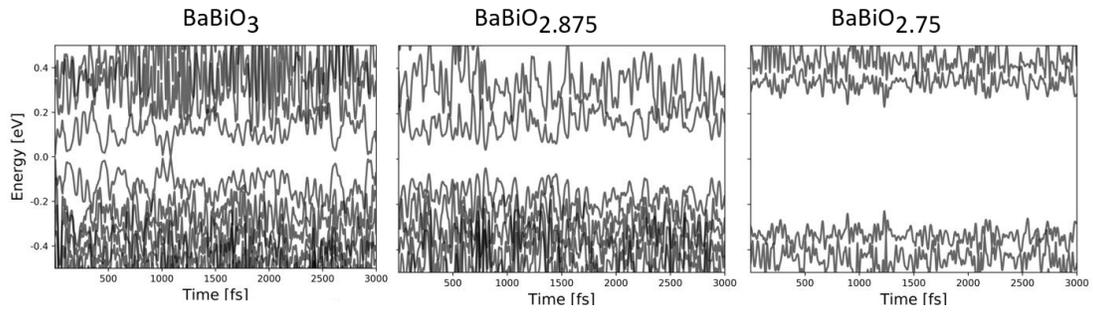

**Supplementary Figure 4** Time evolutions of energy states of BaBiO$_3$, BaBiO$_{2.875}$, BaBiO$_{2.75}$ in MD simulation at 300 K, showing band gap enhancement with electron doping.

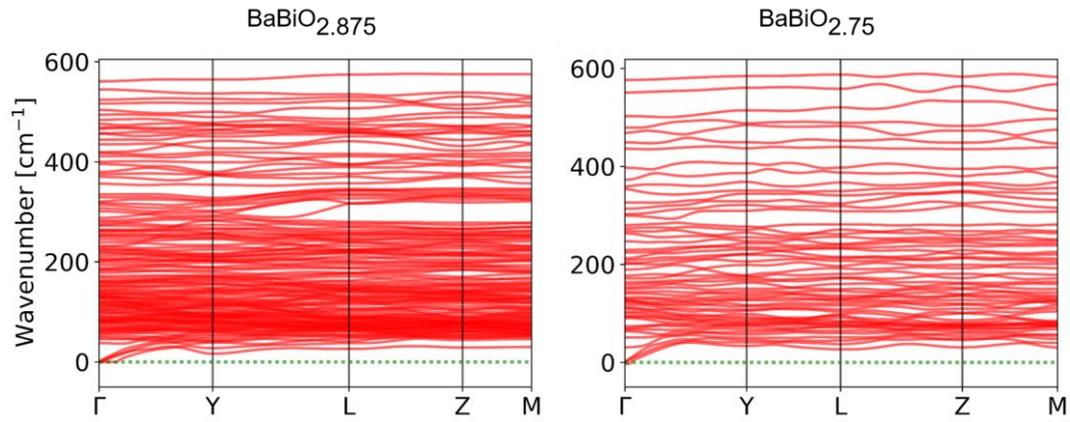

**Supplementary Figure 5** The phonon band structures of BaBiO$_{2.875}$ and BaBiO$_{2.75}$, no negative phonon frequency is found, indicating the structural stability when the oxygen vacancy is introduced.

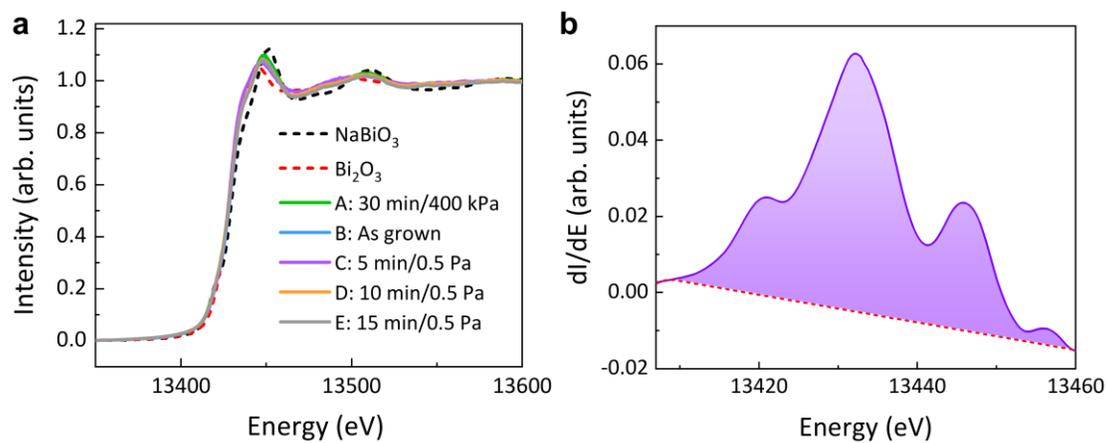

**Supplementary Figure 6** (a) Bi $L_3$-edge XANES spectra of the BaBiO$_{3-\delta}$ ($0 \leq \delta \leq 0.25$) thin films and two reference samples. (b) Integration graph of dI/dE.

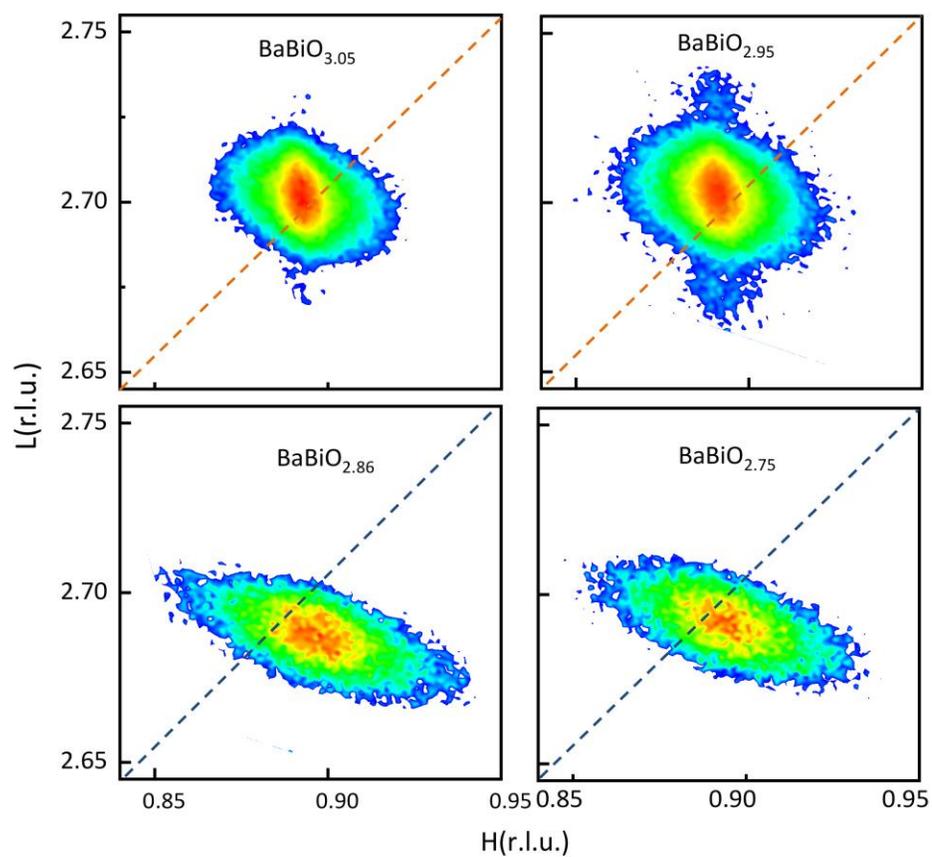

**Supplementary Figure 7** XRD reciprocal space maps of BaBiO$_{3-\delta}$ thin films taken around the (1 0 3) film diffraction peaks; the reciprocal lattice units (r.l.u.) are drawn using the SrTiO$_3$ substrate lattice constants and all thin films are relaxed. The intensity is shown on a logarithmic scale. The dashed lines are guides for eye.

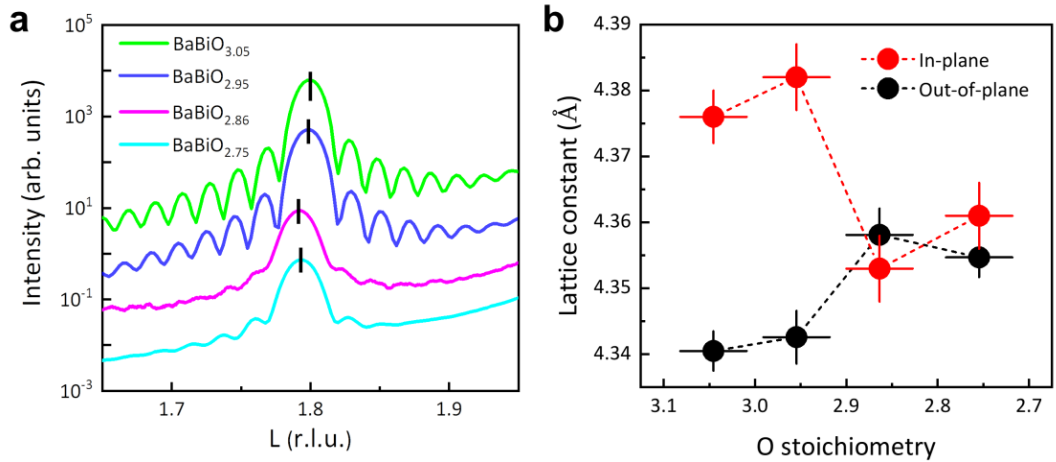

**Supplementary Figure 8** (a) Specular 00L XRD data (H, K=0) for the thin films near L=2, in reciprocal lattice units (r.l.u.) of $SrTiO_3$; the spectra are vertically shifted for clarity. The peaks of thin films are indicated by black lines. The thickness Laue fringes are very pronounced. (b) In-plane and out-of-plane lattice constants of the four sets samples. The dashed lines are guides for eye.

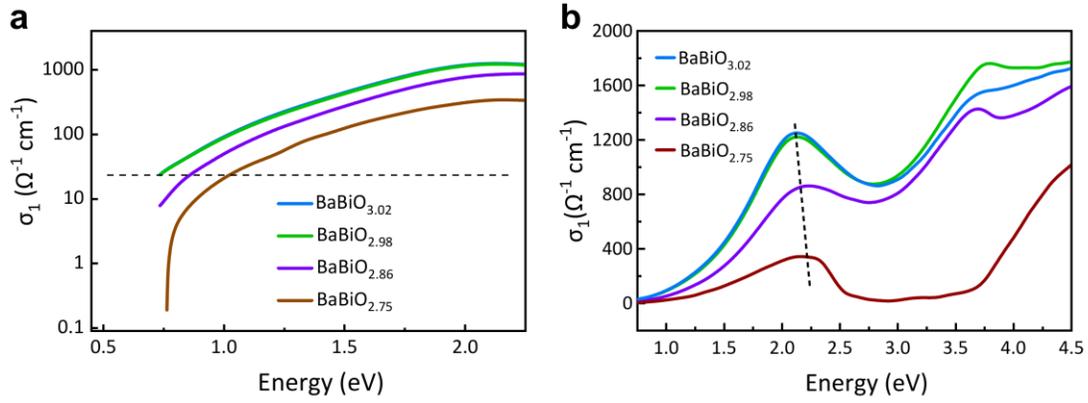

**Supplementary Figure 9** (a) The method to find the photon energy at which the optical conductivity has a very small value of 23 $\Omega^{-1}$ cm$^{-1}$, indicated by the dashed line, in the BaBiO$_{3-\delta}$ ($0 \leq \delta \leq 0.25$) thin films. (b) The large range real part of optical conductivity $\sigma_1$ measured via spectroscopic ellipsometry.

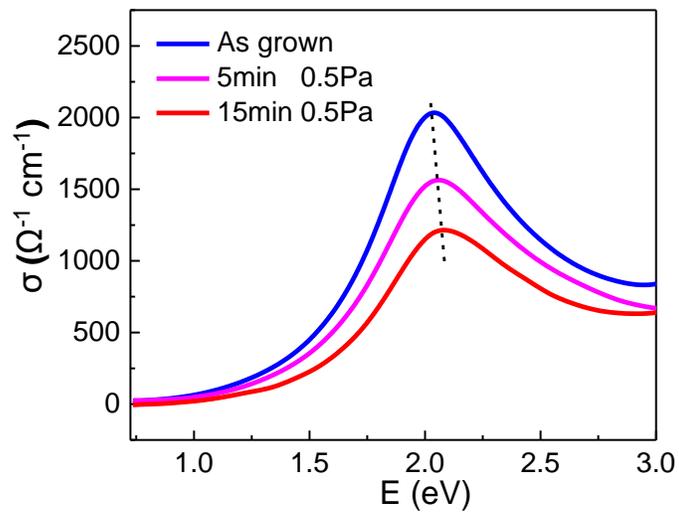

**Supplementary Figure 10** The optical conductivity of another group of samples without any oxygen stoichiometry quantification.

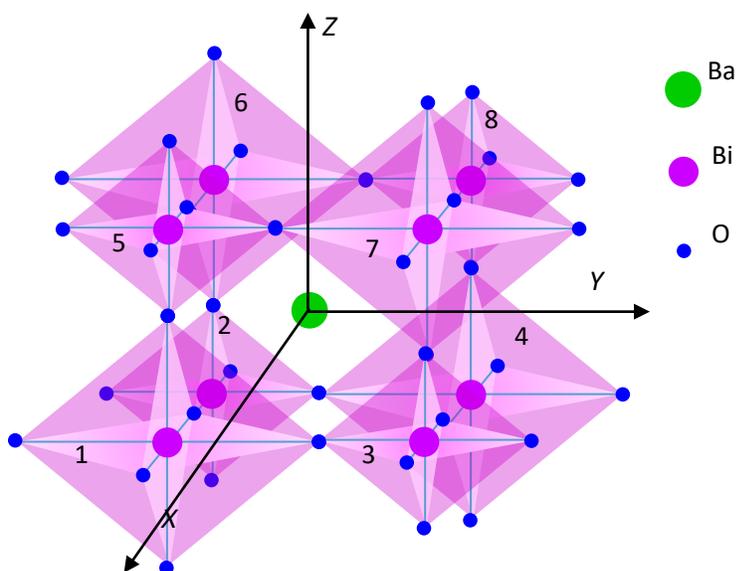

**Supplementary Figure 11** Definition of a coordinate system within a super unit cell of 2x2x2 as a result of breathing distortion.

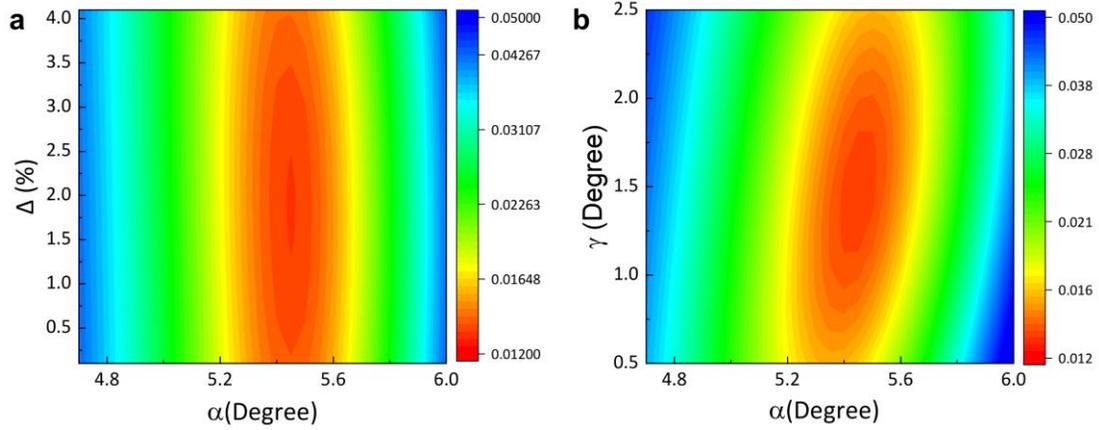

**Supplementary Figure 12 Figure of merit of fitting goodness for oxygen octahedral rotation angles and lattice breathing distortions of BaBiO$_{2.92}$** (a) with $\beta$ and $\gamma$ fixed at the optimal values and (b) with $\beta$ and $\varDelta$ fixed at the optimal values.